\documentclass[preprintnumbers,nofootinbib,noshowpacs,eqsecnum,prd,superscriptaddress,letterpaper]{revtex4}

\usepackage{graphicx}
\usepackage{amsmath,amssymb}
\usepackage{url}
\usepackage{multirow}
\usepackage{hyperref,color}
\graphicspath{{figs/}}
\usepackage{xcolor}
\newcommand{\Frac}[2]{\frac{\displaystyle #1}{\displaystyle #2}}

\newcommand{\sla}[1]{/\!\!\!\!#1}

\newcommand{\cwsq}{{\rm c_W^2} }
\newcommand{\swsq}{{\rm s_W^2} }

\newcommand{\st}{\rm s_W}
\newcommand{\ct}{\rm c_W}
\newcommand{\cdt}{\rm c_{2\theta_W}}

\newcommand{\Dfb}{\mbox{$\raisebox{2mm}{\boldmath ${}^\leftrightarrow$}
\hspace{-4mm} D$}}
\newcommand{\Dfba}{\mbox{$\raisebox{2mm}{\boldmath ${}^\leftrightarrow$}\hspace{-4mm} D^a$}}

\newcommand {\fbw}     {f_{BW}}
\newcommand {\fpone}   {f_{\Phi,1}}

\preprint{\bf YITP-SB-18-14} 

\begin{document}

\title{Effect of Fermionic Operators on the Gauge Legacy of the LHC Run I}
\author{Alexandre Alves}
\affiliation{Departamento de F\'isica, Universidade Federal de S\~ao Paulo, UNIFESP,
Diadema, S\~ao Paulo, Brazil}
\author{N. Rosa-Agostinho}
\affiliation{Departament  de  Fisica  Quantica  i  Astrofisica
 and  Institut  de  Ciencies  del  Cosmos,  Universitat
 de Barcelona, Diagonal 647, E-08028 Barcelona, Spain}
\author{Oscar J. P. \'Eboli}
\affiliation{Instituto de F\'isica, Universidade de S\~ao Paulo, S\~ao Paulo, Brazil}
\author{M.~C.~Gonzalez--Garcia}
\affiliation{C.N. Yang Institute for Theoretical Physics, Stony Brook University, Stony Brook NY11794-3849,  USA}
\affiliation{Departament  de  Fisica  Quantica  i  Astrofisica
 and  Institut  de  Ciencies  del  Cosmos,  Universitat
 de Barcelona, Diagonal 647, E-08028 Barcelona, Spain}
\affiliation{Instituci\'o Catalana de Recerca i Estudis Avancats (ICREA)
Pg. Lluis  Companys  23,  08010 Barcelona, Spain.}
%

\begin{abstract}

  We revisit the extraction of the triple electroweak  gauge boson couplings
  from the
  Large Hadron Collider Run I data on the $W^+W^-$ and $W^\pm Z$
  productions when the analysis also contains additional operators
  that modify the couplings of the gauge bosons to light quarks and
  the gauge boson self-energies. We work in the framework of effective
  Lagrangians where we consider dimension-six operators and perform
  a global fit to consistently take into account the bounds on these
  additional operators originating
  from the electroweak precision data. We show that the constraints on
  the Wilson coefficients $f_B/\Lambda^2$ and $f_W/\Lambda^2$ are
  modified when we include the additional operators while  the
  limits on $f_{WWW}/\Lambda^2$ remain unchanged.
%

\end{abstract}

\maketitle

\section{Introduction}
\label{sec:intro}

The CERN Large Hadron Collider (LHC) has already accumulated an
impressive amount of data that allow for precise tests of the Standard
Model (SM), as well as, a plethora of searches for physics beyond the
standard model. Moreover, the discovery of a new state~\cite{
  Aad:2012tfa, Chatrchyan:2012xdj}, probably the Higgs boson predicted
by the SM, was the first step into the direct exploration of the
electroweak symmetry breaking sector. \smallskip

Within the framework of the SM, the trilinear and quartic vector-boson
couplings are completely determined by the $SU(2)_L \otimes U(1)_Y$
gauge symmetry. Therefore, the scrutiny of these interactions can
either lead to an additional confirmation of the SM or give some hint
on the existence of new phenomena at a higher scale. The triple gauge
couplings (TGC) were for the first time directly probed at
LEP2~\cite{lep2}. At the LHC the largest available center-of-mass
energy allows for further tests of TGC. In fact, the ATLAS and CMS
collaborations studies of the $W^+ W^-$~\cite{Aad:2016wpd,
  Khachatryan:2015sga} and $W^\pm Z$~\cite{Aad:2016ett,
  Khachatryan:2016poo} productions were already used to constrain the
TGC.\smallskip

The combined analysis of the LHC Run I production of electroweak gauge
boson pairs performed in Ref.~\cite{Butter:2016cvz} showed that the
TGC measurement is already dominated by the LHC data with better
precision than the previous results from
LEP2~\cite{lep2}. Furthermore, the analyses of the Higgs boson
properties in the framework of dimension-six effective operators can
indirectly shed light on TGC~\cite{deCampos:1997ez, Corbett:2013pja}
improving the accuracy on the TGC determination~\cite{
  Corbett:2015ksa, Corbett:2012dm, Falkowski:2015jaa,
  Falkowski:2016cxu, deBlas:2017wmn, Ellis:2018gqa}. \smallskip

Recently, Refs.~\cite{Zhang:2016zsp, Baglio:2017bfe} discussed that
changes in the couplings of gauge bosons to fermions, even within the
constraints of electroweak precision data (EWPD), could lead to
modifications of the kinematical distributions in gauge boson pair
production of comparable size to the ones stemming from the purely
anomalous TGC.  This motivate us to revisit the analyses of the LHC
Run I data on the leptonic $W^+W^-$ and $W^\pm Z$ productions to
quantify the impact of anomalous couplings of gauge bosons to fermion
pairs on the TGC bounds when consistently including in the statistical
analysis the EWPD that comprise $Z$ peak observables~\cite{ALEPH:2005ab},
$W$ observables ~\cite{ALEPH:2010aa} and the Higgs
mass~\cite{Aad:2015zhl}.  \smallskip

We work in the framework of effective Lagrangians parametrizing the
departures from the SM by dimension-six operators. Altogether, as
described in Section~\ref{sec:dim6}, the combined analysis of Run I
and EWPD data comprises a total of 11 operators of which a subset of 9
enter the gauge boson pair production at LHC via modifications of the
TGC, of the gauge boson couplings to fermions, as well as,
contributions to the oblique parameters.  Section~\ref{sec:frame}
contains the details of our analyses while our results are presented
in Section~\ref{sec:bounds}. They show that the largest impact on the
LHC Run I constraints on TGC is on the operator ${\cal O}_B$ for which
its 95\%CL allowed range shifts and it also becomes $\sim$ 30\%
wider. The impact on ${\cal O}_W$ is somewhat smaller while the
constraints on ${\cal O}_{WWW}$ are not affected by the inclusion of
the additional operators.  We summarize and discuss our results
in Section~\ref{sec:sum}.

\section{Dimension-six operators}
\label{sec:dim6}

In this work we are interested in deviations from the Standard Model
relevant to gauge boson pair production at the LHC. We parametrize
those in terms of higher dimension operators as
\begin{equation}
   {\cal L}_{\rm eff} = {\cal L}_{\rm SM} + \sum_{n>4,j}
   \frac{f_{n,j}}{\Lambda^{n-4}} {\cal O}_{n,j} \;,
\end{equation}
where the ${\cal O}_{n,j}$ operators are linearly invariant under SM
gauge group $SU(3)_C \otimes SU(2)_L \otimes U(1)_Y$.  Here we assume
$C$ and $P$ conservation.  The first operators that impact the LHC
physics are of $n=6$, {\em i.e.}  dimension--six.  The most general
dimension-six operator basis respecting the SM gauge symmetry, as well
as baryon and lepton number conservation, contains 59 independent
operators, up to flavor and hermitian
conjugation~\cite{Buchmuller:1985jz, Grzadkowski:2010es}.  Since the
S-matrix elements are unchanged by the use of the equations of motion
(EOM), there is a freedom in the choice of basis~\cite{
  Politzer:1980me, Georgi:1991ch, Arzt:1993gz, Simma:1993ky}. Here we
work in that of Hagiwara, Ishihara, Szalapski, and
Zeppenfeld~\cite{Hagiwara:1993ck, Hagiwara:1996kf} for the pure
bosonic operators.

\subsection{Bosonic Operators}

There are nine $C$- and $P$-conserving dimension--six operators in our
basis involving only bosons that take part at tree level in
two--to--two scattering of gauge and Higgs bosons after we employ the
EOM to eliminate redundant operators~\cite{Corbett:2012ja}.  Of those,
five contribute to electroweak gauge boson pair production at LHC
after finite renormalizaton effects are accounted for. In particular
there is just one operator that contains exclusively gauge bosons
\begin{equation}
  \mathcal{O}_{WWW} = {\rm Tr}[\widehat{W}_{\mu}^{\nu}
    \widehat{W}_{\nu}^{\rho}\widehat{W}_{\rho}^{\mu}] \;\;.
\label{eq:www}
\end{equation}
In addition there are four dimension-six operators that include Higgs
and electroweak gauge fields
\begin{equation}
\begin{array}{ll}
 \mathcal{O}_{W}		
=	(D_\mu\Phi)^\dagger\widehat{W}^{\mu\nu}(D_\nu\Phi)  \;\;,	
  &\mathcal{O}_{B}		
=	(D_\mu\Phi)^\dagger\widehat{B}^{\mu\nu}(D_\nu\Phi)  \;\;,
\\
&
\\
\mathcal{O}_{BW}		
= \Phi^\dagger\widehat{B}_{\mu\nu}\widehat{W}^{\mu\nu}\Phi  \;\;,	
  &\mathcal{O}_{\Phi,1}		
=	(D_\mu\Phi)^\dagger\Phi\Phi^\dagger(D^\mu\Phi) \;\;.
\end{array}
\label{eq:phi3}
\end{equation}
Here $\Phi$ stands for the Higgs doublet and $\sigma^a$ are the Pauli
matrices.  We have also adopted the notation $\widehat{B}_{\mu\nu}
\equiv i(g^\prime/2)B_{\mu\nu}$ and $\widehat{W}_{\mu\nu} \equiv
i(g/2)\sigma^aW^a_{\mu\nu}$, with $g$ and $g^\prime$ being the
$SU(2)_L$ and $U(1)_Y$ gauge couplings respectively. \smallskip

The first three operators in Eqs.~(\ref{eq:www})--(\ref{eq:phi3})
modify directly the TGC, thus, affecting the $f \bar{f} \to VV$
scattering, where $V$ stands for the electroweak gauge bosons.  On the
other hand, the operator ${\cal O}_{BW}$ (${\cal O}_{\Phi,1}$) is
associated to the $S$ ($T$) oblique parameter and is constrained by
the EWPD. However, these operators also modify the TGC once the
Lagrangian is canonically normalized.  \smallskip

\subsection{Operators with fermions}

Our operator basis contains 40 independent fermionic operators,
barring flavor indices, that conserve $C$, $P$ and baryon number and
that do not involve gluon fields. Of these operators there are 28 of
them that do not take part in our analyses since they either modify
Higgs couplings to fermions or are contact interactions. Moreover, we
also did not considered 6 that give rise to dipole interactions for
the gauge bosons. Therefore, the operators that contribute to the
processes here analyzed are
\begin{equation}
  \begin{array}{l@{\hspace{1cm}}l@{\hspace{1cm}}l}
& 
   {\cal O}^{(1)}_{\Phi L,ij} =\Phi^\dagger (i\, \Dfb_\mu \Phi) 
(\bar L_{i}\gamma^\mu L_{j}) \;\;,
& 
{\cal O}^{(3)}_{\Phi L,ij}
=\Phi^\dagger (i\,{\Dfba}_{\!\!\mu} \Phi) 
(\bar L_{i}\gamma^\mu T_a L_{j}) \;\;, 
\\
&
&
\\
& 
{\cal O}^{(1)}_{\Phi Q,ij}=
\Phi^\dagger (i\,\Dfb_\mu \Phi)  
(\bar Q_i\gamma^\mu Q_{j}) \;\;,
& 
{\cal O}^{(3)}_{\Phi Q,ij}=\Phi^\dagger (i\,{\Dfba}_{\!\!\mu} \Phi) 
(\bar Q_i\gamma^\mu T_a Q_j) \;\;,
\\
&
&
\\
& {\cal O}^{(1)}_{\Phi u,ij}=\Phi^\dagger (i\,\Dfb_\mu \Phi) 
(\bar u_{R_i}\gamma^\mu u_{R_j}) \;\;,
& {\cal O}^{(1)}_{\Phi d,ij}=\Phi^\dagger (i\,\Dfb_\mu \Phi) 
(\bar d_{R_i}\gamma^\mu d_{R_j}) \;\;,
\\
&
&
\\
& 
{\cal O}^{(1)}_{\Phi ud,ij}=\tilde\Phi^\dagger (i\,\Dfb_\mu \Phi) 
(\bar u_{R_i}\gamma^\mu d_{R_j} +{\rm h.c.}) \;\;,
& 
{\cal O}^{(1)}_{\Phi e,ij}=\Phi^\dagger (i\Dfb_\mu \Phi) 
(\bar e_{R_i}\gamma^\mu e_{R_j})  \;\;,
\end{array}
\label{eq:hffop}
\end{equation}
where we defined $\tilde \Phi=i \sigma_2\Phi^*$,
$\Phi^\dagger\Dfb_\mu\Phi= \Phi^\dagger D_\mu\Phi-(D_\mu\Phi)^\dagger
\Phi$ and
$\Phi^\dagger \Dfba_{\!\!\mu} \Phi= \Phi^\dagger T^a D_\mu
\Phi-(D_\mu\Phi)^\dagger T^a \Phi$ with $T^a=\sigma^a/2$.  We have
also used the notation of $L$ for the lepton doublet, $Q$ for the
quark doublet and $f_R$ for the $SU(2)_L$ singlet fermions, where
$i, j$ are family indices. \smallskip

In order to avoid the existence of blind
directions~\cite{DeRujula:1991ufe, Elias-Miro:2013mua} in the analyses
of the EWPD we used the freedom associated to the use of EOM to remove
from our basis the following combination of
operators~\cite{Corbett:2012ja}
\begin{equation}
  \sum_i {\cal O}^{(1)}_{\Phi L,ii} \;\;, \;\;\;{\rm and}\;\;\;
  \sum_i {\cal O}^{(3)}_{\Phi L,ii} \;\;.
\label{eq:EOMred}  
\end{equation}
Furthermore, to prevent the generation of too large flavor violation,
in what follows we assume no generation mixing in the above operators.
For the same reason we will work under the assumption that the
coefficient of the potential source of additional flavour violation,
${\cal O}^{(1)}_{\Phi ud,ij}$, is suppressed and can be neglected
\footnote{This operator contributes only to the right-handed coupling
  of the $W$, therefore it does not interfere with the SM amplitudes
  and, at linear order, is not constrained by the EWPD.}.  Also for
simplicity we consider the operators to be generation independent.  In
this case operators ${\cal O}^{(1)}_{\Phi L}$ and
${\cal O}^{(3)}_{\Phi L}$ are removed by the use of EOM.  \smallskip
Therefore, in our basis, only the operator
${\cal O}^{(1)}_{\Phi e,ij}$ modifies the $Z$ coupling to leptons,
while there are additional contributions to the $Z$-quark pair
vertices originating from $ {\cal O}^{(1)}_{\Phi u,ij}$,
$ {\cal O}^{(1)}_{\Phi d,ij}$, ${\cal O}^{(1)}_{\Phi Q,ij}$, and
${\cal O}^{(3)}_{\Phi Q,ij}$.  Moreover, the $W$ coupling to fermions
receives extra contributions from ${\cal O}^{(3)}_{\Phi Q,ij}$ and
${\cal O}^{(1)}_{\Phi ud,ij}$.  \smallskip

Operators $\mathcal{O}_{BW}$, $\mathcal{O}_{\Phi,1}$,
${\cal O}^{(1)}_{\Phi Q}$, ${\cal O}^{(3)}_{\Phi Q}$,
${\cal O}^{(1)}_{\Phi Q}$, ${\cal O}^{(1)}_{\Phi u}$,
${\cal O}^{(1)}_{\Phi d}$, and ${\cal O}^{(1)}_{\Phi e}$ can be
bounded by the EWPD, in particular from $Z$--pole and $W$--pole
observables~\cite{Corbett:2017qgl}.  In this work we focus on
fermionic operators most relevant for gauge boson pair production at
LHC which are those leading to modification of the quark couplings to
gauge bosons and we will not consider ${\cal O}^{(1)}_{\Phi e}$ in our
TGC analysis but it is kept in the EWPD studies. Furthermore, for
completeness, we also include the effect of the dimension--six
four--fermion operator contributing with a finite renormalization to
the Fermi constant
\begin{equation}
  {\cal O}_{LLLL}=(\bar L \gamma^\mu L)(\bar L \gamma^\mu L) \;\;.
\end{equation}

Altogether the effective Lagrangian considered in this work reads:
\begin{eqnarray}
{\cal L}_{eff} = {\cal L}_{SM} 
&+& \frac{f_{WWW}}{\Lambda^2} {\cal  O}_{WWW}
+ \frac{f_{W}}{\Lambda^2} {\cal O}_{W}
+ \frac{f_{B}}{\Lambda^2} {\cal O}_{B}
+ \frac{f_{BW}}{\Lambda^2} {\cal O}_{BW}
+ \frac{f_{\Phi,1}}{\Lambda^2} {\cal O}_{\Phi,1}
\nonumber
\\
&+& \frac{f^{(1)}_{\Phi Q}}{\Lambda^2}   {\cal O}^{(1)}_{\Phi Q}
  +     \frac{f^{(3)}_{\Phi Q}}{\Lambda^2}   {\cal O}^{(3)}_{\Phi Q}
  +     \frac{f^{(1)}_{\Phi u}}{\Lambda^2}   {\cal O}^{(1)}_{\Phi u}
  +     \frac{f^{(1)}_{\Phi d}}{\Lambda^2}   {\cal O}^{(1)}_{\Phi d}
  +     \frac{f^{(1)}_{\Phi e}}{\Lambda^2}   {\cal O}^{(1)}_{\Phi e} 
  +     \frac{f_{LLLL}}{\Lambda^2}   {\cal O}_{LLLL} \;\;. 
\label{eq:leff}
\end{eqnarray}

\smallskip

\subsection{Lorentz and $U(1)_{em}$ invariant  Parametrization}

After accounting for finite renormalization effects, the part of
Lagrangian (\ref{eq:leff}) relevant for our analyses can be cast in a
Lorentz and $U(1)_{em}$ invariant form as:
\begin{align}
\Delta {\cal L}_{f,V}\equiv &
- i e \; \Delta \kappa_\gamma \; W^+_\mu W^-_\nu \gamma^{\mu \nu}
- \frac{i e \lambda_\gamma}{2 M_W^2} \; W_{\mu \nu}^+ W^{- \nu \rho} \gamma_\rho^{{}\;\mu}
- \frac{i e \ct \lambda_Z}{2 M_W^2} \; W_{\mu \nu}^+ W^{- \nu \rho} Z_\rho^{\;\mu} 
\notag \\
& -  i e \ct \; \Delta \kappa_Z \; W^+_\mu W^-_\nu Z^{\mu \nu} - i e \ct \; \Delta g_1^Z \;
\left( W^+_{\mu \nu} W^{- \mu} Z^\nu - W^+_\mu Z_\nu W^{- \mu \nu} 
\right)
\notag
\\
&
-\frac{e}{\st \ct} Z_\mu \sum_{f} \bar\psi^f\gamma^\mu
\left[ \Delta g^f_L P_L+\Delta g^f_R P_R\right]\psi^f
-\frac{e}{\sqrt{2}\st} \left[ W^+_\mu 
  \left(\bar \psi^u\gamma^\mu \Delta g^{ud}_{WL} P_L \psi^d +
  \bar \psi^\nu\gamma^\mu \Delta g^{e\nu}_{WL} P_L \psi^e\right)+
  h.c.\right]
\;\;,
\end{align}
where $P_{L,R}$ are the chirality projectors. The TGC effective
couplings are
\begin{equation}
\begin{array}{l@{\hspace{1.2cm}}ll}  
\Delta \kappa_\gamma = \Frac{e^2 v^2}{8\swsq\Lambda^2}
  \left( f_W + f_B  -2 f_{BW}\right) \;\;,
&&
 \Delta g_1^Z = \Frac{e^2 v^2}{8 \swsq\cwsq\Lambda^2} \left(f_W
+\Frac{2\swsq}{\cdt}
f_{BW}\right)-\Frac{1}{4\cdt}\frac{v^2}{\Lambda^2} 
\fpone \;\;,
\\[+0.5cm]
    \Delta \kappa_Z =\Delta g_1^Z-\frac{\swsq}{\cwsq} \Delta\kappa_\gamma \;\;,
&&
\lambda_\gamma = \lambda_Z = 
\Frac{3 e^2 M_W^2}{2\swsq \Lambda^2} f_{WWW}\;\;. \\ 
\end{array}
\end{equation}
where $c_W$ ($s_W$) stands for the cosine (sine) of the weak mixing
angle and $\cdt$ is the cosine of twice this angle.
Notice that there are only three independent TGC due to the linear
realization of the $SU(2)_L \otimes U(1)_Y$ symmetry in the
dimension-six operators. The effective couplings of the fermions can
be written as
\begin{equation}
\begin{array}{l@{\hspace{1.2cm}}ll}  
\Delta g_{L,R}^f=g_{L,R}^f\Delta g_1+Q^f\Delta g_{2}+\Delta\tilde g^f_{L,R}\,. 
&&\Delta g_{WL}^{ff'}=\Delta g_W+ \Delta\tilde g_{WL}^{ff'}\;\; ,
\end{array}
\end{equation}
where $g_L^f=T_3^f-\swsq Q^f$ and $g_R^f=-\swsq Q^f$ are the SM
couplings. The first contributions to these anomalous couplings
originates from finite renormalizations due to ${\cal O}_{BW}$ and
${\cal O}_{\Phi,1}$
\begin{equation}
\begin{array}{l@{\hspace{1.2cm}}l@{\hspace{1.2cm}}l}  
\Delta g_1=\Frac{1}{2}\left(\alpha\, T\right) \;\;,
&
\Delta g_2=\Frac{\st^2}{\cdt}\left(\ct^2 \left(\alpha\, T \right) -
\frac{1}{4\st^2}\alpha\, S\right)\;\; , &
\Delta g_W=\Frac{\ct^2}{2\cdt} \alpha\, T -\Frac{1}{4\cdt} \alpha\, S
\;\;,
\end{array}
\end{equation}
with the oblique parameters given by
\begin{equation}
\begin{array}{l@{\hspace{1.2cm}}l@{\hspace{1.2cm}}l}  
  \alpha\,S=-e^2\Frac{v^2}{\Lambda^2}\fbw \;\;,
&
  \alpha\, T= -\Frac{1}{2}\Frac{v^2}{\Lambda^2}\fpone \;\; . 
\end{array}
\end{equation}
The fermionic dimension-six operators in Eq.\ (\ref{eq:leff})
give rise to additional contributions
\begin{equation}
\begin{array}{l@{\hspace{1.2cm}}l}  
\Delta \tilde g^u_{L}=
-\frac{v^2}{8 \Lambda^2} (4f^{(1)}_{\Phi Q}- f^{(3)}_{\Phi Q}) \;\; , 
\qquad\qquad
&
\Delta \tilde g^u_{R}= -\Frac{v^2}{2 \Lambda^2}  f^{(1)}_{\Phi u} \;\; , 
\\[+0.5cm] 
\Delta \tilde g^d_{L}= -\Frac{v^2}{8 \Lambda^2}
(4f^{(1)}_{\Phi Q}+ f^{(3)}_{\Phi Q}) \;\; ,
&
\Delta \tilde g^d_{R}= -\Frac{v^2}{2 \Lambda^2}  f^{(1)}_{\Phi d} \;\;,
\\[+0.5cm] 
\Delta\tilde g_{WL}^{ud}=\Frac{v^2}{4 \Lambda^2}  f^{(3)}_{\Phi Q} \;\;,
& \Delta \tilde g^e_{R}=-\frac{v^2}{2 \Lambda^2}  f^{(1)}_{\Phi e} \;\;.
\label{eq:effcoupl}
\end{array}
\end{equation}

The anomalous TGC and gauge bosons interactions to quarks modify the
high energy behavior of the scattering of quark pairs into two
electroweak gauge bosons since the anomalous interactions can spoil
the cancellations built in the SM. For the $W^+W^-$ and $W^\pm Z$
channels the leading scattering amplitudes in the helicity basis are
\begin{eqnarray}
&& A( d_- \bar{d}_+ \to W^+_0 W^-_0)  = i \frac{s}{\Lambda^2} \sin
   \theta \left \{ - \frac{g^2}{24 c_W^2} (3 c_W^2 f_W - s_W^2 f_B ) +
   \frac{1}{4} ( f^{(3)}_{\Phi Q } - 4 f^{(1)}_{\Phi Q} )
\right \} \;\;,
\label{eq:grow1}
\\
&& A( d_- \bar{d}_+ \to W^+_\pm W^-_\pm)  = - i \frac{s}{\Lambda^2} \sin 
   \theta \frac{3  g^4}{8} f_{WWW}  \;\;,
\label{eq:grow2}\\
&& A( d_+ \bar{d}_- \to W^+_0 W^-_0)  = - i \frac{s}{\Lambda^2} \sin 
   \theta \left \{ \frac{g^2 s_W^2}{12 c_W^2} f_B  + f^{(1)}_{\Phi d} 
\right \} \;\;,
\label{eq:grow3}\\
&& A( u_- \bar{u}_+ \to W^+_0 W^-_0)  = i \frac{s}{\Lambda^2} \sin
   \theta \left \{ \frac{g^2}{24 c_W^2} (3 c_W^2 f_W + s_W^2 f_B ) -
   \frac{1}{4} ( f^{(3)}_{\Phi Q } + 4 f^{(1)}_{\Phi Q} )
\right \} \;\;,
\label{eq:grow4}\\
&& A( u_+ \bar{u}_- \to W^+_0 W^-_0)  = i \frac{s}{\Lambda^2} \sin 
   \theta \left \{ \frac{g^2 s_W^2}{6 c_W^2} f_B  - f^{(1)}_{\Phi u} 
\right \} \;\;,
\label{eq:grow5}\\
&& 
A( u_- \bar{u}_+ \to W^+_\pm W^-_\pm)  =  
A( d_- \bar{u}_+ \to Z_\pm W^-_\pm)  =  
i \frac{s}{\Lambda^2} \sin    \theta \frac{3 g^4 }{8} f_{WWW}  \;\;,
\label{eq:grow6}\\
&&
A( d_- \bar{u}_+ \to W^-_0 Z_0)= i \frac{s}{\Lambda^2} \sin    \theta \left \{
\frac{g^2}{4 \sqrt{2}} f_W - \frac{1}{2 \sqrt{2}} f^{(3)}_{\Phi Q}
\right \} \;\;,
\label{eq:grow7}
\end{eqnarray}
where $s$ stands for the center-of-mass energy and $\theta$ is the
polar angle in the center-of-mass frame.\smallskip

We notice that the leptonic operator ${\cal O}^{(1)}_{\Phi e}$ does
not contribute to the gauge boson production amplitudes at LHC.  It
only contributes to the decay rate of the $Z$ boson in the $ZW$
channels and in the narrow width approximation its effect is
subdominant. We will not consider it in the TGC analysis but it is
kept in the EWPD analysis. \smallskip

\section{ANALYSES FRAMEWORK}
\label{sec:frame}

In order to constrain the parameters in the effective Lagrangian in
Eq.~\eqref{eq:leff} we study the $W^+W^-$ and $W^\pm Z$ productions in
the leptonic channel since these are the measurements with the highest
sensitivity for charged triple gauge boson vertices. In doing so we
consider the same kinematic distributions employed by the experiments
for their anomalous gauge boson coupling analyses what allows us to
validate our results against the bounds obtained by the experiments in
each of the final states. More specifically, the channels that we
analyze and their kinematical distributions are \smallskip
\begin{center} \begin{tabular}{l|lcr|c@{\extracolsep{0.25cm}}
        c@{\extracolsep{0.25cm}}c}
\hline 
Channel ($a$) & Distribution & \# bins ($N_{b}$)  & Data set &
$\sigma_{\rm sig}$ 
& $\sigma_{\rm bck}$  & $\sigma_{i,\rm unc}$ 
\\ [0mm]
\hline
$WW\rightarrow \ell^+\ell^{\prime -}+\sla{E}_T\; (0j)$
& $p^{\rm leading, lepton}_{T}$
& 3 & ATLAS 8 TeV, 20.3 fb$^{-1}$~\cite{Aad:2016wpd}
& 0.049   & 0.02& 0.08 -- 0.14 
\\[0mm]
$WW\rightarrow \ell^+\ell^{(\prime) -}+\sla{E}_T\; (0j)$
& $m_{\ell\ell^{(\prime)}}$ & 8 & CMS 8 TeV, 19.4 fb$^{-1}$~\cite{Khachatryan:2015sga}
& 0.069 & 0.02 & 0.01 -- 0.08
\\[0mm]
$WZ\rightarrow \ell^+\ell^{-}\ell^{(\prime)\pm}$ & $m_{T}^{WZ}$ & 6 & ATLAS 8 TeV, 20.3 fb$^{-1}$~\cite{Aad:2016ett}
& 0.1 & 0.02 & 0.12 -- 0.18
\\[0mm]
$WZ\rightarrow \ell^+\ell^{-}\ell^{(\prime)\pm}+\sla{E}_T$ & $Z$ candidate $p_{T}^{\ell\ell}$ & 10 & CMS 8 TeV, 19.6 fb$^{-1}$~\cite{Khachatryan:2016poo}
& 0.15 & 0.02 & 0.15 -- 0.25
\\[0mm]
\hline
  \end{tabular}
\end{center}\smallskip
For each experiment and channel, we extract from the experimental
publications the observed event rates in each bin, $N^{a}_{i,\rm d}$,
as well as the background expectations $N^{a}_{i,\rm bck}$, and the SM
$W^+W^-$ ($W^\pm Z$) predictions, $N^{a}_{i,\rm sm}$. \smallskip

The procedure to obtain the relevant kinematical distributions
predicted by Eq.~\eqref{eq:leff} is as follows.  First we simulate the
$W^+W^-$ and $W^\pm Z$ productions using
\textsc{MadGraph5}~\cite{Alwall:2014hca} with the UFO files for our
effective Lagrangian generated with
\textsc{FeynRules}~\cite{Christensen:2008py, Alloul:2013bka}.  We
employ \textsc{PYTHIA6.4}~\cite{Sjostrand:2006za} to perform the
parton shower, while the fast detector simulation is carried out with
\textsc{Delphes}~\cite{deFavereau:2013fsa}. In order to account for
higher order corrections and additional detector effects we simulate
SM $W^+W^-$ and $W^\pm Z$ productions in the fiducial region requiring
the same cuts and isolation criteria adopted by the corresponding
ATLAS and CMS studies, and normalize our results bin by bin to the
experimental collaboration predictions for the kinematical
distributions under consideration.  Then we apply these correction
factors to our simulated $WV$ distributions in the presence of the
anomalous couplings.  This procedure yields our predicted number of
signal events in each bin $i$ for the ``$a$'' channel, $N_{i,\rm
  sig}^{a,\rm nosys}$.  \smallskip

The statistical confrontation of these predictions with the LHC Run I
data is made by means of a binned log-likelihood function based on the
contents of the different bins in the relevant kinematical
distribution of each channel. Depending on the number of data events
in the bin we use a Poissonian or a Gaussian probability distribution
for its statistical error.  In constructing the log-likelihood
function we simulate the effect of the systematic and theoretical
uncertainties by introducing two sets of pulls: two globally affecting
the predictions of the event rates in all bins in fully correlated
form -- which parametrize, among others, the luminosity uncertainty,
and theoretical errors on the total cross-section for the process and
its backgrounds-- and $N^a_b$ independent pulls, one per-bin, to
account for the bin-uncorrelated errors arising from the theoretical
errors affecting the distributions, experimental energy resolutions
and, in general, any energy and/or momentum dependence of the
uncertainties. With this, the number of predicted events in bin $i$
for channel $a$ is
$N^a_i=\left[(1+\xi^a_{\rm sig})(1+\xi^a_{i,\rm unc})N^{a,\rm
    nosys}_{i,\rm sig} +(1+\xi^a_{\rm bck})N^{a}_{i,\rm bck} \right]$.
The errors of these pulls are introduced as Gaussian bias in the
log-likelihood functions and are extracted from the information given
by the experiments. For completness they are reported in the table
above.  \smallskip

\begin{figure}[h!]
    \centering
\includegraphics[width=0.7\textwidth]{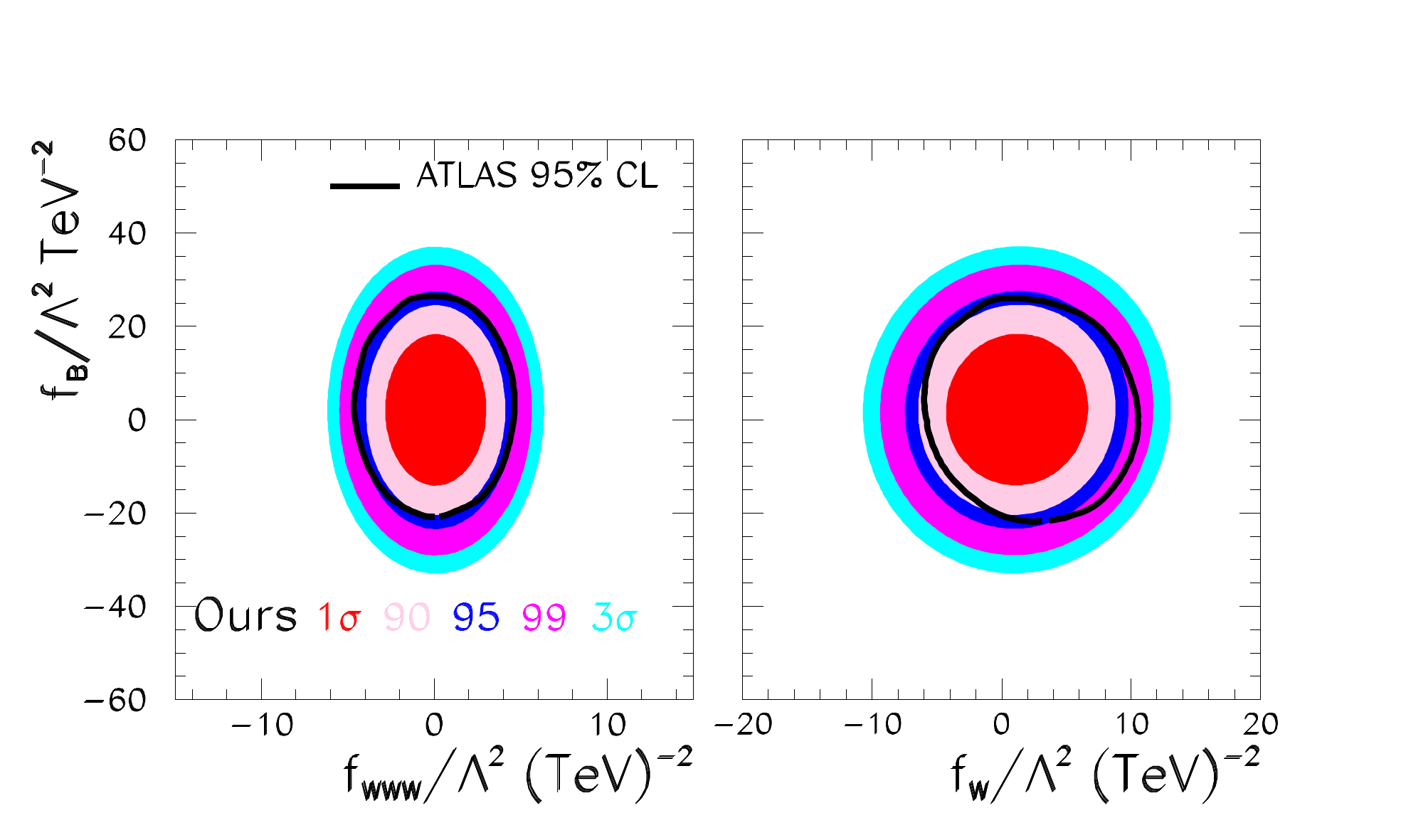}
\caption{ Allowed regions in the planes
  $f_B/\Lambda^2 \otimes f_{WWW}/\Lambda^2$ (left panel) and
  $f_B/\Lambda^2 \otimes f_{W}/\Lambda^2$ (right panel) at $1\sigma$,
  95\%, 99\%, and $3\sigma$ CL. The black line stands for the border
  of the 95\% CL allowed region obtained by ATLAS~\cite{Aad:2016wpd}.
}
  \label{fig:validation}
\end{figure}

In order to validate our simulation we obtain first the 95\% CL
allowed regions for the TGC for each channel and experiment under the
same assumptions the collaboration used. For example, we present in
Figure~\ref{fig:validation} our two-dimensional allowed regions using
the ATLAS $W^+W^-$ data and assuming that the only non-vanishing
Wilson coefficients are $f_{WWW}$, $f_{W}$ and $f_{B}$, two different
from zero at a time, as in the ATLAS analysis.  As seen in the figure,
our results for the 95\% CL allowed region (blue region) agrees well
with the one obtained by ATLAS, whose border is represented by the
black curve. \smallskip

When including the effect of the additional operators we must also
account for their contribution to the EWPD. 
Our construction of the $\chi^2$ function for the EWPD follows the analysis 
in Ref.~\cite{Corbett:2017qgl} to which we refer the reader for details.
In brief in our EWPD analysis we fit 15 observables of which 12 are $Z$
observables~\cite{ALEPH:2005ab}:
\begin{equation*}
\Gamma_Z \;\;,\;\;
\sigma_{h}^{0} \;\;,\;\;
{\cal A}_{\ell}(\tau^{\rm pol}) \;\;,\;\;
R^0_\ell \;\;,\;\;
{\cal A}_{\ell}({\rm SLD}) \;\;,\;\;
A_{\rm FB}^{0,l} \;\;,\;\;
R^0_c \;\;,\;\;
 R^0_b \;\;,\;\;
{\cal  A}_{c} \;\;,\;\;
 {\cal A}_{b} \;\;,\;\;
A_{\rm FB}^{0,c}\;\;,\;\;
\hbox{ and} \;\;
A_{\rm FB}^{0,b}  \hbox{ (SLD/LEP-I)}\;\;\; ,
\end{equation*}
complemented by three $W$ observables
\begin{equation*}
  M_W   \;\;,\;\; \Gamma_W \;\;\hbox{  and}\;\; \hbox{Br}( W\to {\ell\nu})
\end{equation*}
that are, respectively, its average mass taken
from~\cite{Olive:2016xmw}, its width from
LEP2/Tevatron~\cite{ALEPH:2010aa}, and the leptonic $W$ branching
ratio for which the average in Ref.~\cite{Olive:2016xmw} is
considered.  The correlations among these inputs are presented in
Ref.~\cite{ALEPH:2005ab} and we take them into consideration in the
analyses. The SM predictions and their uncertainties due to variations of
the SM parameters were extracted from~\cite{Ciuchini:2014dea}.
\smallskip

Altogether we construct a combined $\chi^2$ function
\begin{eqnarray}
  \chi^2_{\rm LHC-RI\, +\, EWPD}&\equiv&\chi^2_{\rm LHC-RI}
    (f_W,f_B,f_{WWW},f_{BW},f_{\Phi,1},f^{(1)}_{\phi,Q},f^{(3)}_{\phi,Q},
  f^{(1)}_{\phi,u},f^{(1)}_{\phi,d}) \nonumber \\& +& 
  \chi^2_{\rm EWPD}(f_{BW},f_{\Phi,1},f^{(1)}_{\phi,Q},f^{(3)}_{\phi,Q},
  f^{(1)}_{\phi,u},f^{(1)}_{\phi,d},f^{(1)}_{\phi,e},f_{LLLL}) \; ,
\label{eq:chiglo}
\end{eqnarray}
from which we derive the allowed ranges for each coefficient or pair
of coefficients after marginalization over all the others. 
It is worth commenting that this marginalization of
the profiled binned log-likelihood is computationally very expensive
due the high dimensionality of the parameters space. Achieving an
acceptable accuracy in the determination of the statistical confidence
bounds for one- and two-dimensional distributions requires typically
hundreds of millions of Monte Carlo evaluations for each one of the
points used to obtain the 1D and 2D the allowed
regions.\smallskip

Finally, for comparison, we also consider the constraints from LEP2
global analysis of TGC~\cite{lep2}. In order to do so we follow the
procedure in Ref.~\cite{Butter:2016cvz} and construct a simplified
gaussian $\chi^2_{\rm LEP2}$ using the central values, $\sigma$ and
correlation matrix for the couplings $\Delta g^1_Z$,
$\Delta \kappa_\gamma$ and $\lambda$ and their correlation
coefficients from the final combined LEP2 analysis in Ref.~\cite{lep2}
(reproduced in Table~\ref{tab:lep2} for completness) which was
performed in terms of these effective TGC coefficients under the
relations implied by dimension-six effective operator formalism for
TGC. We notice, however, that in extracting those bounds on the
effective TGC couplings, the LEP collaborations did not include the
effect of fermion operators. For that reason the combination of those
LEP2 bounds with our LHC Run I and EWPD is only shown for the purpose
of illustration.  \smallskip

\begin{table}[t]
\begin{tabular}{c|crrr}
\hline 
&  \multicolumn{4}{c}{LEP} \\ 
& \phantom{xxxxxx} 68 \% CL \phantom{xxxxxx} & \multicolumn{3}{c}{Correlations} \\ \hline
$\Delta g_1^Z$ & \phantom{$-$} $0.051^{+0.031}_{-0.032}$ & $1.00$ & $0.23$ & $-0.30$ \\[1mm]
$\Delta \kappa_\gamma$ & $-0.067^{+0.061}_{-0.057}$ & $0.23$ & $1.00$ & $-0.27$ \\[1mm]
$\lambda$ & $-0.067^{+0.036}_{-0.038}$ & $-0.30$ & $0.27$ & $1.00$ \\
\hline
\end{tabular}
\caption {$\Delta g_1^Z$, $\Delta \kappa_\gamma$ and  $\lambda$
 central values, standard deviations and correlation
  coefficients from LEP2~\cite{lep2}.}
\label{tab:lep2}
\end{table}

\section{BOUNDS ON TRIPLE GAUGE BOSON INTERACTIONS}
\label{sec:bounds}
  \begin{figure}[h!]
    \centering
\includegraphics[width=\textwidth]{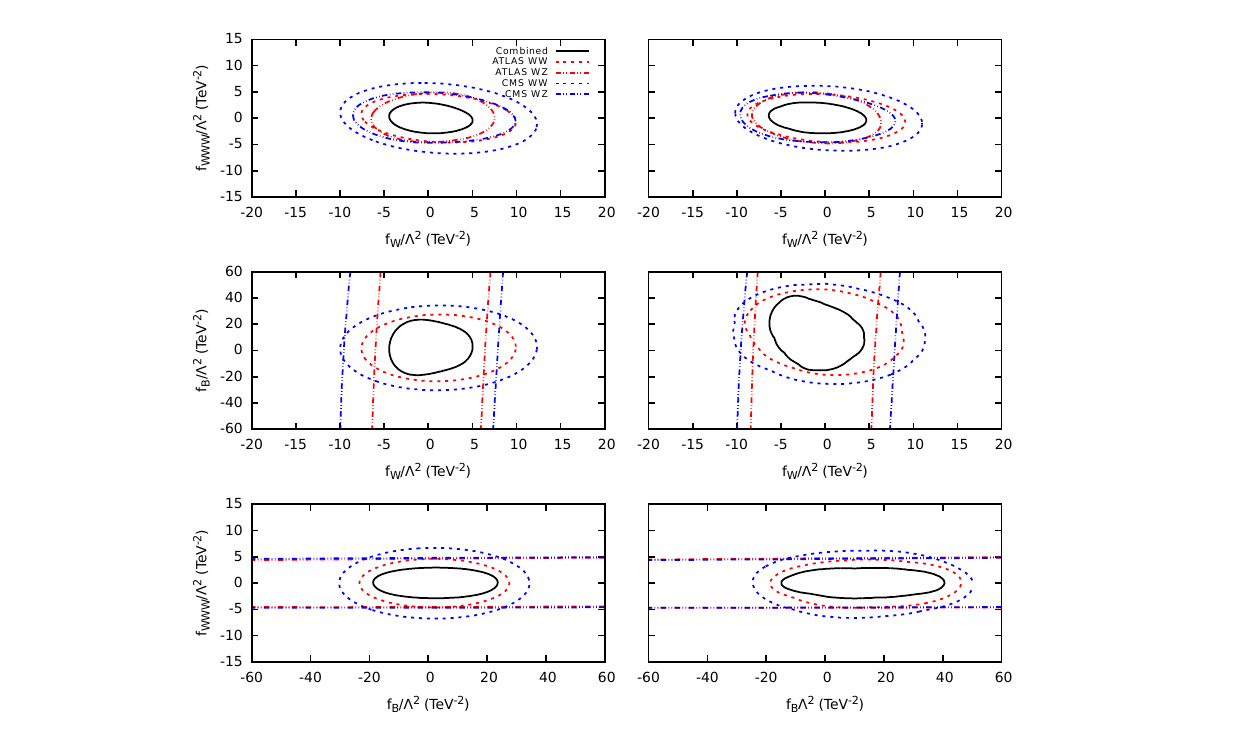}
\caption{ Allowed 95\% CL regions in the planes
  $f_{WWW}/\Lambda^2 \otimes f_{B}/\Lambda^2$ (top row),
  $f_B/\Lambda^2 \otimes f_{W}/\Lambda^2$ (middle row)  and
  $f_{WWW}/\Lambda^2 \otimes f_{W}/\Lambda^2$ (lower row) for the
  different channels as labeled in the figure. In the left panels only
  $f_{WWW}$, $f_{W}$ and $f_{B}$ were considered non-zero in the fit,
  while the right panels display the result from the 11 parameter fit.
  In each case we marginalize over the undisplayed non-zero
  variables.}
  \label{fig:combined}
\end{figure}

We start by showing the results of our analysis in terms of the
allowed ranges for the Wilson coefficients of the three ``canonical''
TGC operators, $f_{WWW}/\Lambda^2$, $f_{W}/\Lambda^2$ and
$f_{B}/\Lambda^2$. We depict first in Fig.~\ref{fig:combined} the 95\%
CL (2 dof) allowed regions in the planes
$f_B/\Lambda^2 \otimes f_{W}/\Lambda^2$,
$f_{WWW}/\Lambda^2 \otimes f_B /\Lambda^2$ and
$f_{WWW}/\Lambda^2 \otimes f_{W}/\Lambda^2$ for the $W^+W^-$ and
$W^\pm Z$ channels and for ATLAS and CMS, as well as the combination
of these results. In order to assess the impact of additional
operators in the TGC extraction at LHC we performed first the
``standard'' analysis fitting just these three coefficients, and
setting the coefficient of all other operators to zero. The
corresponding allowed regions are shown in the left panels after
marginalizing over the third coefficient which is not
displayed. Conversely the results of the global analysis of the LHC
Run I data together with EWPD performed in terms of 11 non-zero Wilson
coefficients (see Eq.~\eqref{eq:leff}) are shown on the right panels.
These regions are obtained after marginalization over the 9
undisplayed coefficients. \smallskip
  \begin{figure}[h!]
    \centering
\includegraphics[width=0.5\textwidth]{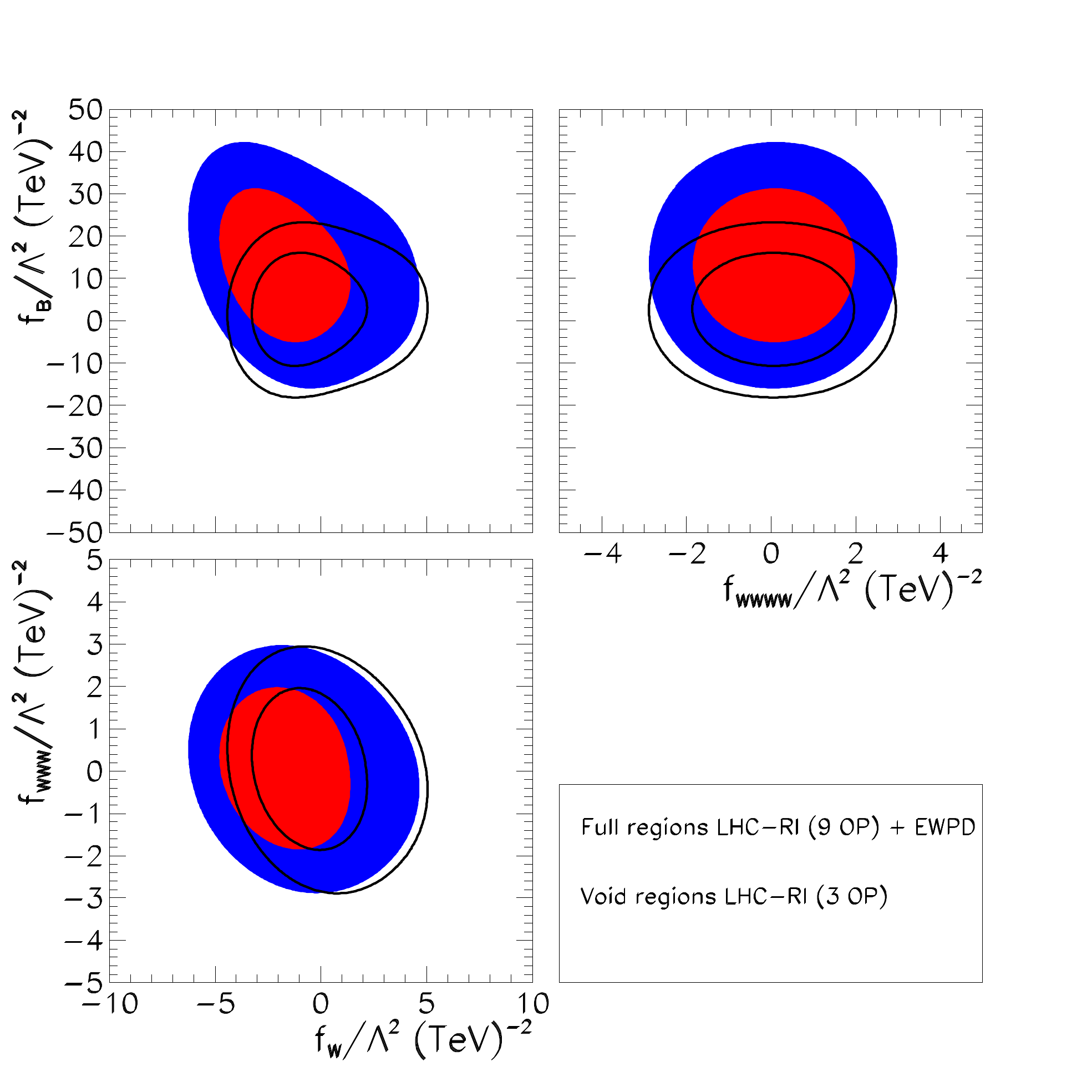}
\caption{ $1\sigma$ and 95\% CL allowed regions in the planes
  indicated in the axes. Here we considered the $W^+W^-$ and $W^\pm Z$
  productions and the EWPD in the analyses. The data set and
  parameters used are as indicated in the figure.  }
  \label{fig:2dbos}
\end{figure}

One salient feature of Figure~\ref{fig:combined} is that the $W^\pm Z$
bounds on the Wilson coefficient $f_B/\Lambda^2$ are much looser than
the ones on $f_W/\Lambda^2$ and $f_{WWW}/\Lambda^2$ , as expected,
because ${\cal O}_B$ does not contribute to the leading term of the
growth of the scattering amplitudes; see
Eqs.~\eqref{eq:grow1}--\eqref{eq:grow7}. \smallskip

For better comparison of the results obtained with and without
including the additional operators we overlay in Fig.~\ref{fig:2dbos}
the $1\sigma$ and 95\% CL allowed regions obtained combining all
channels and experiments for the two scenarios.  As we can see from
this figure the addition of more parameters leads to the expansion of
the allowed regions, as expected. Moreover, the region of
$f_B/\Lambda^2$ suffers the largest shift towards positive values of
this parameter while there is a small shift in the $f_W/\Lambda^2$
direction and there is no appreciable displacement along the
$f_{WWW}/\Lambda^2$ axis. \smallskip

  \begin{figure}[h!]
    \centering
\includegraphics[width=0.7\textwidth]{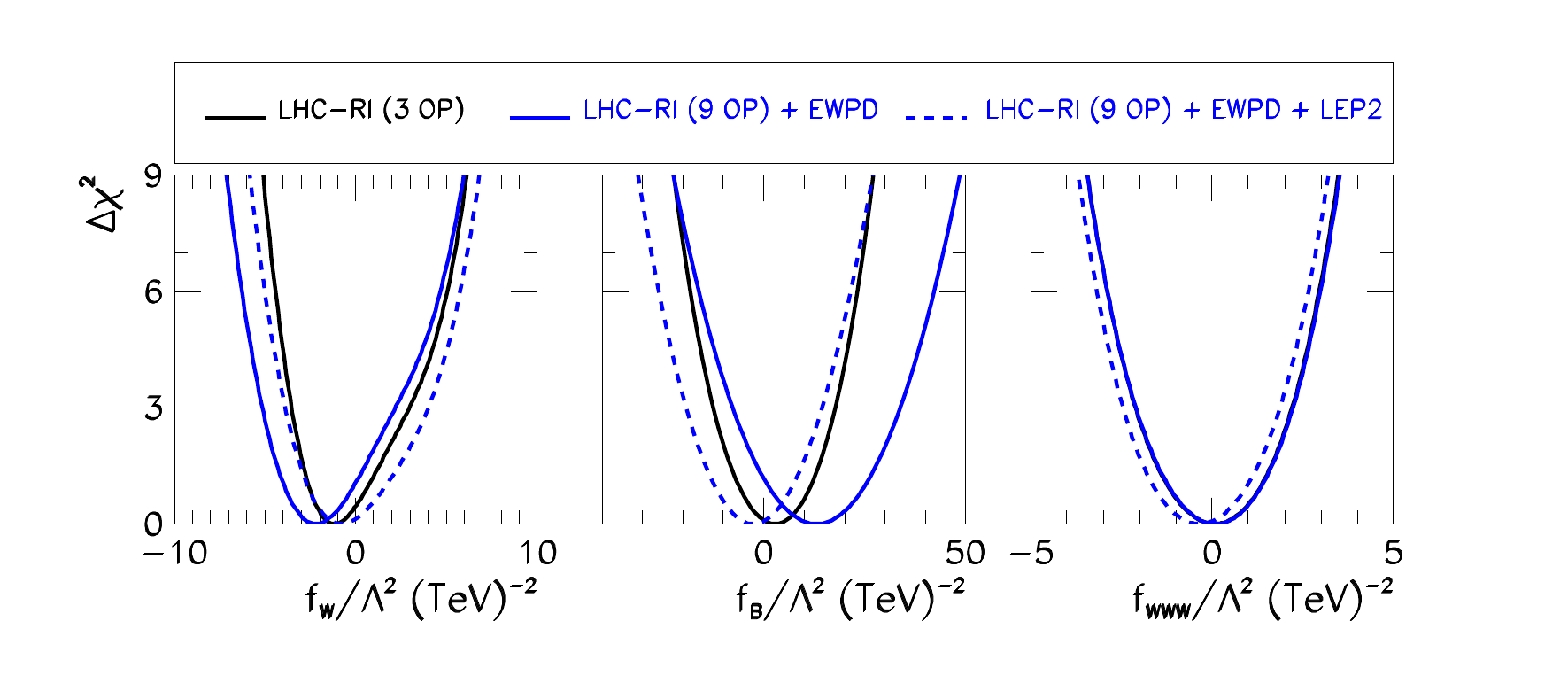}
\caption{ $\Delta \chi^2$ dependence on the $f_W/\Lambda^2$ (left
  panel), $f_{B}/\Lambda^2$ (central panel) and $f_W/\Lambda^2$ (right
  panel) parameters after the marginalization over the remaining fit
  parameters. The solid black line stands for the standard TGC analysis,
  while the solid blue line represents the 9 parameter fit
to the LHC Run I data and the EWPD. The dashed blue line differs from
the solid ones just by the addition of LEP2 data on TGC.}
  \label{fig:1dtgc}
\end{figure}

The corresponding dependence of the $\Delta\chi^2$ for the two
analysis with each of the three coefficients is given in
Fig.~\ref{fig:1dtgc} and from those we read the 95\%CL one-dimensional
allowed ranges for each coefficient given in Table~\ref{tab:ewpd}.  As
seen above, the $\Delta \chi^2$ distribution for $f_B/\Lambda^2$
($f_W/\Lambda^2$) broadens and shifts to positive (negative) values
when we compare the results considering only the LHC Run I data and
three canonical parameters (solid black line) with the one containing
additional operators also constrained by the EWPD (solid blue
line). Quantitatively the effect is slightly larger for
$f_B/\Lambda^2$ whose allowed range widens by about 30\% versus 20\%
for $f_W/\Lambda^2$. \smallskip
  \begin{figure}[h!]
    \centering
\includegraphics[width=0.8\textwidth]{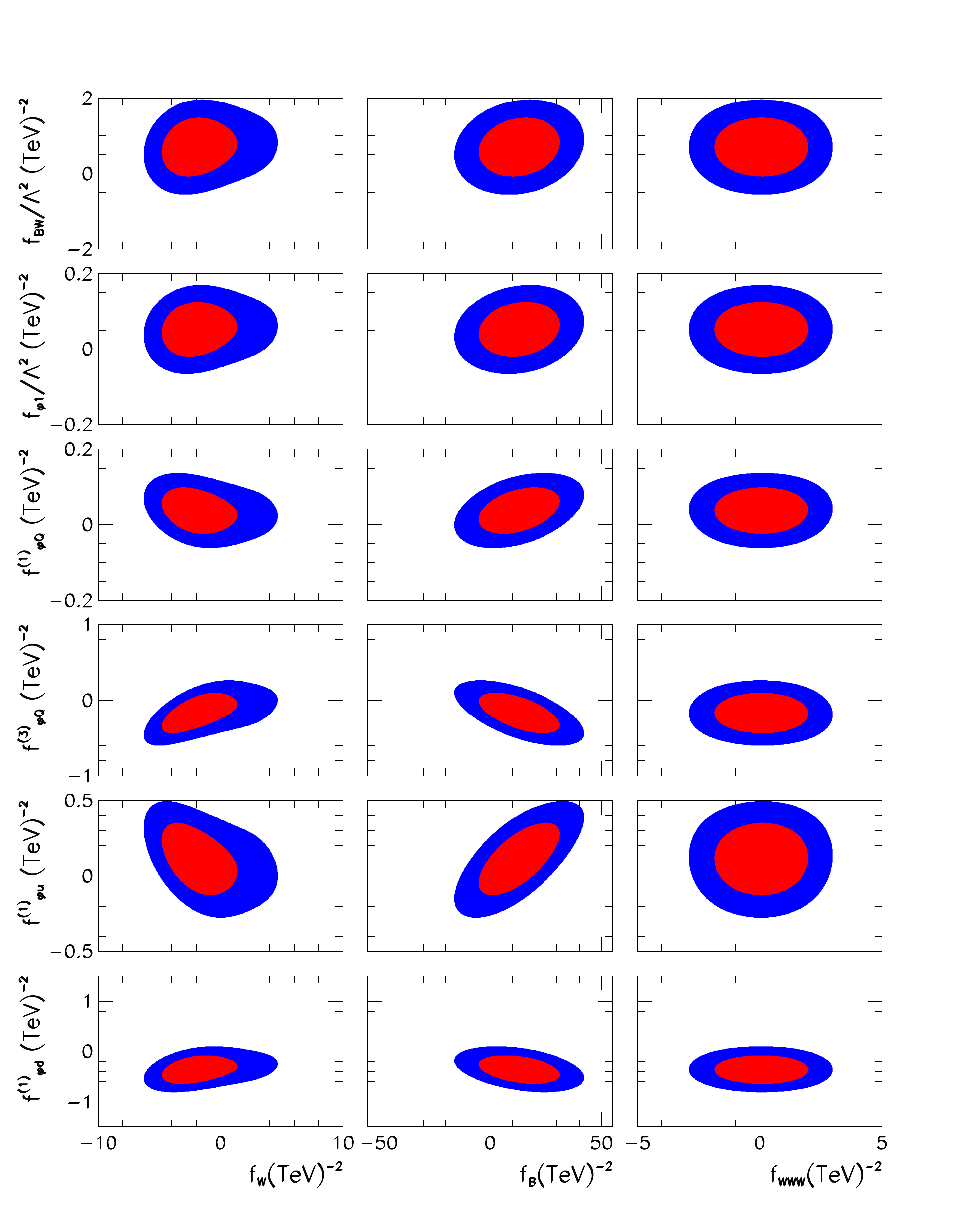}
\caption{ $1\sigma$ and 95\% CL allowed regions in the planes
  indicated in the axes. Here we considered the $W^+W^-$ and $W^\pm Z$
  productions and the EWPD in the analyses.}
  \label{fig:bosvsferm}
\end{figure}

The effect of each of the six additional operators on the extracted
range of the three ``canonical'' TGC operator coefficients is
illustrated in Figure~\ref{fig:bosvsferm} where we depict the
two-dimensional correlations between the three TGC coefficients and
the additional ones.  In each panel of this figure we exhibit the
$1\sigma$ and 95\% CL level (2 dof) allowed regions after
marginalizing over the remaining parameters.
As we can see, $f_B/\Lambda^2$ has a significant correlation only with
$f^{(1)}_{\Phi u}/\Lambda^2$
and to a lesser extent is (anti-)
correlated with $f^{(1)}_{\Phi Q}/\Lambda^2$
($f^{(3)}_{\Phi Q}/\Lambda^2$ and $f^{(1)}_{\Phi d}/\Lambda^2$).
This is expected as these are the operator coefficients contributing
the growth of the scattering amplitudes into longitudinally polarized
gauge bosons (Eqs.~\ref{eq:grow1}--~\ref{eq:grow7}). In particular 
the correlion with $f^{(1)}_{\Phi u}/\Lambda^2$ can be understood from the
scattering amplitude in Eq.~\eqref{eq:grow5}.
Similarly $f_W/\Lambda^2$ shows a stronger anti-correlation only with
$f^{(1)}_{\Phi u}/\Lambda^2$  and to a smaller degree is correlated
with $f^{(3)}_{\Phi Q}/\Lambda^2$ and $f^{(1)}_{\Phi d}/\Lambda^2$.
Finally from the third column of this figure we can see that
$f_{WWW}/\Lambda^2$ shows no correlation with the additional parameters
as expected since ${\cal O}_{WWW}$ contributes by itself to the energy
growth of the scattering amplitudes for transversely polarized gauge
bosons.  \smallskip

\begin{figure}[h!]
    \centering
\includegraphics[width=0.7\textwidth]{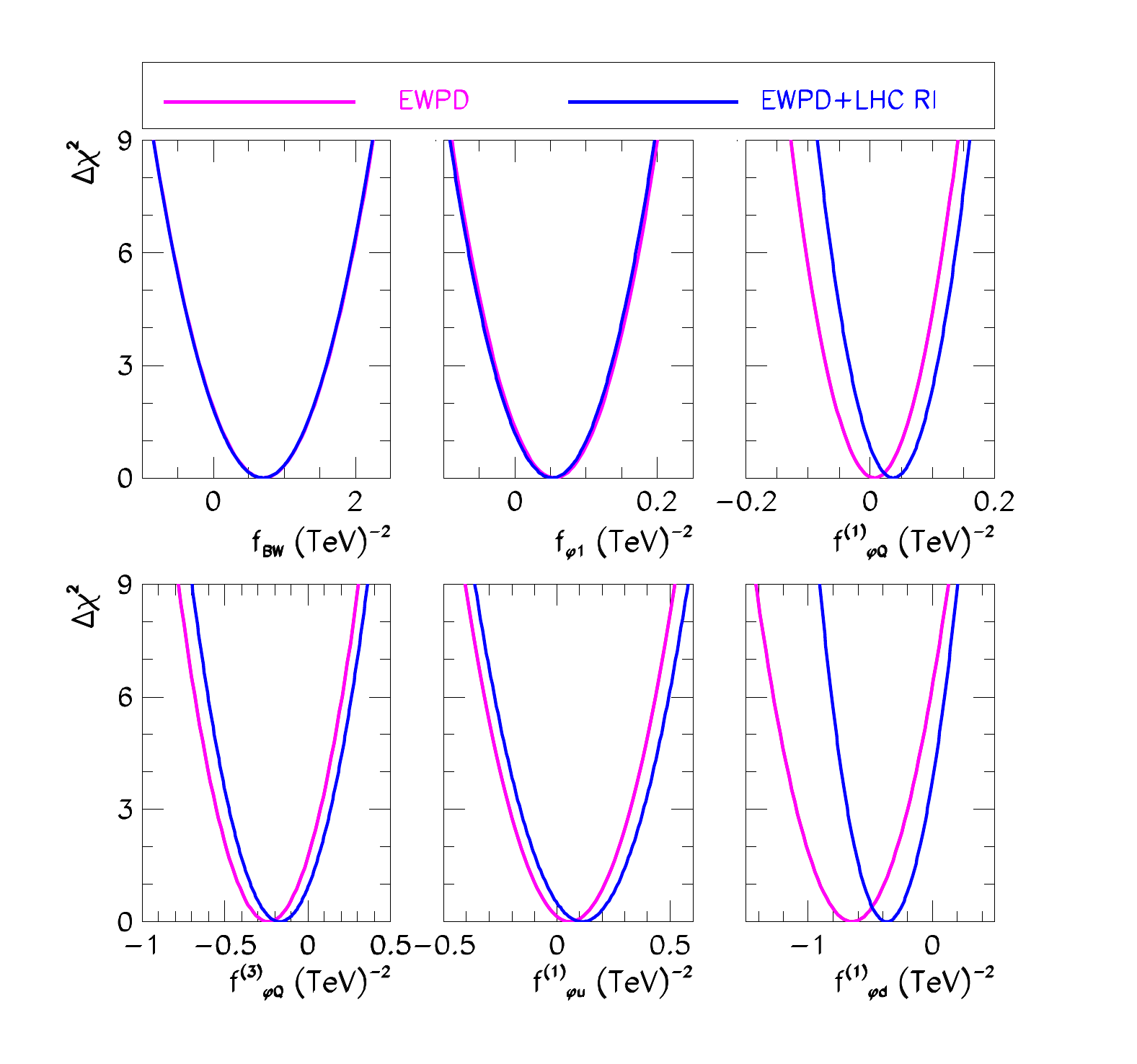}
\caption{ $\Delta \chi^2$ dependence on the $f_{BW}/\Lambda^2$,
  $f_{\Phi 1}/\Lambda^2$, $f^{(1)}_{\Phi Q}/\Lambda^2$,
  $f^{(3)}_{\Phi Q}/\Lambda^2$, $f^{(1)}_{\Phi u}/\Lambda^2$, and
  $f^{(1)}_{\Phi d}/\Lambda^2$ after the marginalization over the
  undisplayed parameters.  The magenta line stands for the results
  using only the EWPD while the blue one is obtained considering the
  EWPD and LHC Run I diboson production.}
  \label{fig:1dferm}
\end{figure}

The impact of the LHC diboson production data on the determination of
the parameters directly constrained by the EWPD is illustrated in
Fig.~\ref{fig:1dferm} that depicts the $\Delta \chi^2$ distribution as
a function of these parameters where the magenta (blue) line stands
for the result obtained using the EWPD (and the LHC Run I diboson
production data).
%

The top left and middle panels of this figure show that the addition
of the LHC data does not alter the constraints on $f_{BW}/\Lambda^2$
and $f_{\Phi 1}/\Lambda^2$ parameters. This is easy to understand
since these parameters do not modify the high energy behavior of
$q \bar{q} \to V V$ amplitudes; see
Eqs.~\eqref{eq:grow1}--\eqref{eq:grow2}. This is expected from
${\cal O}_{\Phi,1}$ as it only contributes to the amplitudes via
finite renormalization effects of the SM parameters.  The operator
${\cal O}_{BW}$, on the other hand, modifies the TGC directly also,
however, its effects on the $Z$ wave-function renormalization cancel
the growth with the center--of--mass energy due to the anomalous TGC.
From the top right, bottom left and middle panels we can see that the
impact of the Run I data on $f^{(1)}_{\Phi Q}/\Lambda^2$,
$f^{(3)}_{\Phi Q}/\Lambda^2$ and $f^{(1)}_{\Phi u}/\Lambda^2$ is
marginal.  $f^{(1)}_{\Phi d}/\Lambda^2$ is the only parameter whose
$\Delta \chi^2$ distribution gets significantly affected. The EWPD
analysis favours non-vanishing value for $f^{(1)}_{\Phi d}/\Lambda^2$
at 2$\sigma$, a result driven by the 2.7$\sigma$ discrepancy between
the observed $A_{\rm FB}^{0,b}$ and the SM. On the contrary no
significant discrepancy is observed between the observed LHC Run I
diboson data and the SM. Hence there is a shift towards zero of
$f^{(1)}_{\Phi d}/\Lambda^2$ when including the LHC Run I data in the
analysis. This slight tension results also into the reduction of the
globally allowed range. \smallskip

  \begin{figure}[h!]
    \centering
\includegraphics[width=0.5\textwidth]{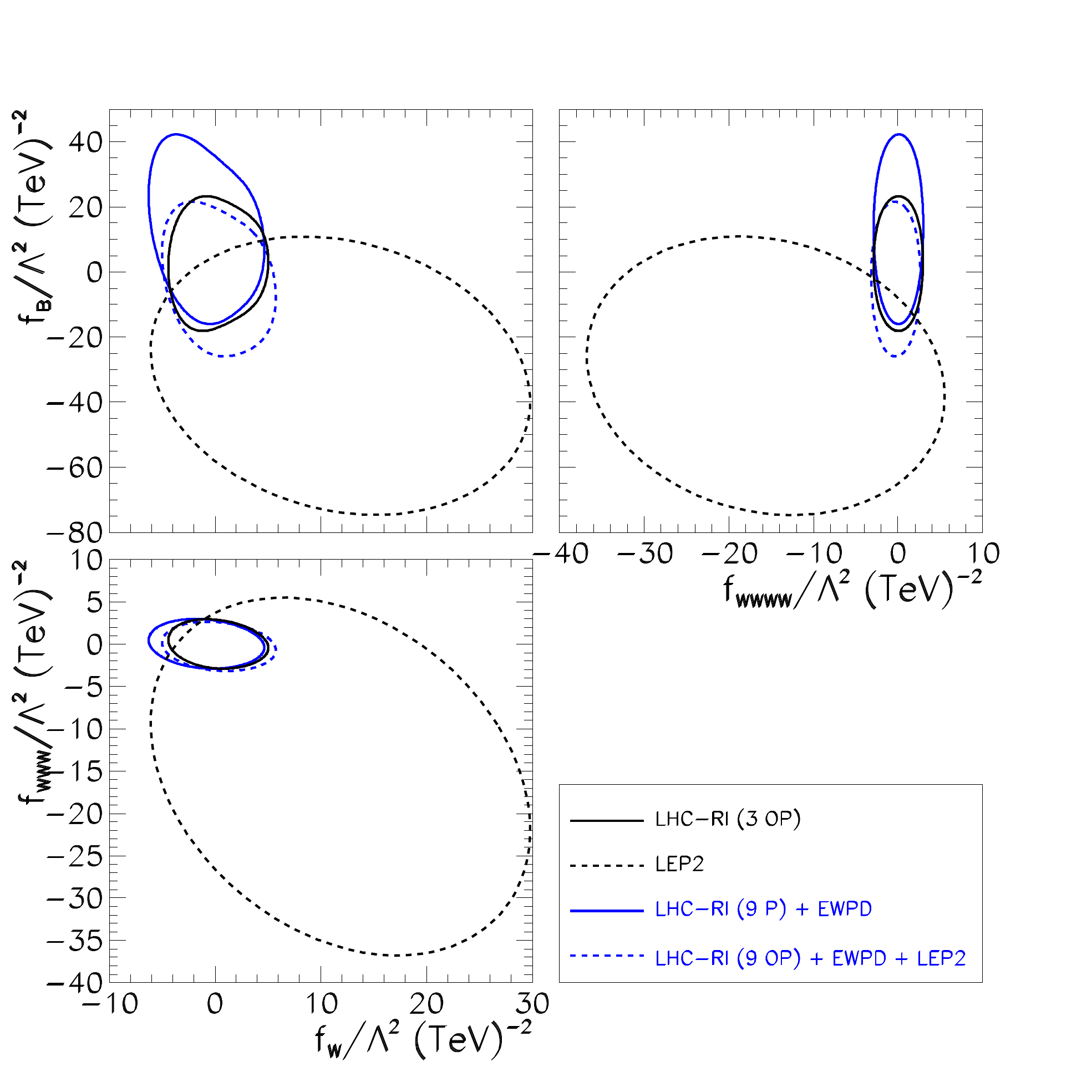}
\caption{ Two-dimensional 95\% CL allowed regions in the planes
  indicated in the axes. Here we considered the $W^+W^-$ and $W^\pm Z$
  productions, LEP2 data on TGC and the EWPD in the analyses. The
  lines are as shown in the figure.}
  \label{fig:2dwithlep}
\end{figure}

We finish this section by comparing our results with the bounds
derived from LEP2 diboson data. To do so we plot in
Figure~\ref{fig:2dwithlep} the two-dimensional 95\% CL allowed regions
for the three combination of the canonical TGC parameters for the
analysis with and without additional operators together with the LEP2
results.  As shown in Ref.\cite{Butter:2016cvz}, the limits emanating
from the canonical LHC Run I diboson data (black solid line) are
substantially more stringent than those imposed by LEP2 (black dashed
line). As seen in this figure, enlarging the number of operators in
the LHC analyses, together with the EWPD, does not alter this
conclusion despite the growth of the allowed regions (solid blue
line). For illustration we also show in the figure the allowed regions
obtained by naively combining the general LHC Run I + EWPD analysis
with the LEP2 information (see discussion at the end of
Sec.~\ref{sec:frame}).  As seen, including  LEP2 data in the
approximation used leads to a
reduction of the allowed regions in the $f_B/\Lambda^2$ direction, as
well as to a shift of it towards negative values (see also the dashed
blue line in Fig.~\ref{fig:1dtgc}). \smallskip

\section{Summary}
\label{sec:sum}

In this work we have quantified the impact of possible anomalous gauge
couplings to quarks on the TGC determination performed using the LHC Run I
diboson data. In order to carry out a statistically consistent analysis we have
included in addition the EWPD to constrain the couplings between quarks and
gauge boson as well as the modifications of the gauge boson self energies.
We have worked in the framework of effective Lagrangians so 
our study has been performed including the 11 dimension-six operators given
in Eq.~\eqref{eq:leff}. \smallskip

As a summary of our findings we present in Table~\ref{tab:ewpd} the
95\% CL globally allowed ranges for the Wilson coefficients of the
nine operators that contribute to the LHC Run I data considered. The
comparison of the the first and third columns of this table shows that
the addition of the new operators modifies the TGC bounds on
$f_B/\Lambda^2$ and $f_W/\Lambda^2$ coming from the LHC Run I diboson.
Quantitatively the effect is slightly larger for $f_B/\Lambda^2$ whose
allowed range widens by about 30\% versus 20\% for
$f_W/\Lambda^2$. The limits on $f_{WWW}/\Lambda^2$, on the contrary,
result almost unaffected.  Despite these changes, the constraints on
these parameters are still dominated by the LHC Run I data at large,
and are still substantially stronger than those obtained from
LEP2 data. \smallskip

We have also learned from our analyses that the LHC Run I diboson data is not
precise enough to yield substantial information on the gauge couplings
to quarks in addition to what is already known from EWPD; contrast the
second and third columns of Table~\ref{tab:ewpd}. The only apparent
exception is $f^{(1)}_{\Phi d} / \Lambda^2$ which in the considered
family universal scenario is driven to be non-zero in the EWPD
analysis by the discrepancy between the measured $A_{\rm FB}^{0,b}$ at
LEP/SLC and the SM while LHC Run I data shows no evidence of any
deviation with respect to the SM.  Nevertheless, these results allow
us to foresee that diboson production at the LHC will play an important
role in the analyses of anomalous couplings of gauge bosons to quarks
as the LHC increases the integrated luminosity. Hence global analysis
of LHC and EWPD are becoming a must for consistent determination of
the Wilson coefficients of the full set of dimension-6 operators.\smallskip

\begin{table}[h!]\centering
\renewcommand{\arraystretch}{1.2}
\begin{tabular}{|c|c|c|c|}
\hline 
coupling & \multicolumn{3}{c|}{95\% allowed range  (TeV$^{-2})$} \\
\hline 
&  LHC RI (3 OP)   & EWPD   &  LHC RI (9 OP) + EWPD  \\
\hline\hline
$f_W$& $(-3.9\,,\,3.9)$ 
&---
&$(-5.6\,,\,4.0)$ 
\\
$f_B$& $(-15\,,\,20)$ 
&---
& $(-11\,,\,37)$ 
\\
$f_{WWW}$& $(-2.4\,,\,2.5)$ 
&---
& $(-2.4\,,\,2.6)$ 
\\
$f_{BW}$ & ---
&  $(-0.32\,,\,1.7)$ 
& $(-0.33\,,\,1.7)$ 
\\
$f_{\Phi 1}$ & ---
& $(-0.040\,,\, 0.15)$
&$(-0.044\,,\, 0.15)$
\\
$f^{(1)}_{\Phi Q}$& ---
& $(-0.083  \,,\,0.10   )$
&  $(-0.044  \,,\,0.12   )$
\\
$f^{(3)}_{\Phi Q}$& ---
& $(-0.60  \,,\, 0.12  )$
& $(-0.52  \,,\, 0.18  )$
\\
$f^{(1)}_{\Phi u}$  & ---
&$(-0.25  \,,\, 0.37  )$
& $(-0.19  \,,\, 0.42  )$
\\
$f^{(1)}_{\Phi d}$& ---
& $(-1.2  \,,\, -0.13  )$
&$(-0.73  \,,\, 0.023  )$
\\
\hline
\end{tabular}
\caption{ 95\% CL allowed ranges for the Wilson coefficients of the
  dimension--six operators that contribute to  the studied processes in
  gauge boson pair production at LHC. The ranges for each parameter
  are obtained after marginalization of the coefficients of all other
  operators contributing to each analysis. In particular the results
  given in the third and forth column are obtain after  marginalization
  over $f^{(1)}_{\phi,e}$ and $f_{LLLL}$ as well.}
\label{tab:ewpd}
\end{table}

\acknowledgments 

We thank Tyler Corbett for discussions.  O.J.P.E. and N.R.A. want to
thank the group at SUNY at Stony Brook for the hospitality during the
final stages of this work. This work is supported in part by Conselho
Nacional de Desenvolvimento Cient\'{\i}fico e Tecnol\'ogico (CNPq) and
by Funda\c{c}\~ao de Amparo \`a Pesquisa do Estado de S\~ao Paulo
(FAPESP) grants 2012/10095-7 and <2017/06109-5, by USA-NSF grant
PHY-1620628, by EU Networks FP10 ITN ELUSIVES
(H2020-MSCA-ITN-2015-674896) and INVISIBLES-PLUS
(H2020-MSCA-RISE-2015-690575), by MINECO grant FPA2016-76005-C2-1-P
and by Maria de Maetzu program grant MDM-2014-0367 of ICCUB.  A. Alves
thanks Conselho Nacional de Desenvolvimento Cient\'{\i}fico (CNPq) for
its financial support, grant 307265/2017-0. We are specially indebted 
to Juan Gonzalez Fraile for providing us
with his personal codes and detailed results relevant to the analysis
in Ref.~\cite{Butter:2016cvz}.


\bibliography{references}
\end{document}